\documentclass[onecolumn,showpacs,preprintnumbers,amsmath,amssymb]{revtex4}
\usepackage{graphicx}
\usepackage{dcolumn}
\usepackage{bm}
\usepackage{color}
\usepackage{amsmath}
\usepackage{xcolor}
\usepackage{caption}
\usepackage{float}
\newcommand{\Dmq}{\Delta m^2}
\newcommand{\eVq}{\ensuremath{\text{eV}^2}}

\begin{document}

\title{A novel technique to achieve atomic macro-coherence as a tool to determine 
the nature of neutrinos}

\author{R. Boyero\,$^{1, 3}$}\email{robertobg@usal.es}
\author{A. V. Carpentier\,$^3$} \email{avazquez@clpu.es}
\author{J.\,J.\,Gomez-Cadenas\,$^2$}\email{gomez@mail.cern.ch}
\author{A. Peralta Conde\,$^3$}\email{aperalta@clpu.es}

\affiliation{$^1$ Facultad de Ciencias, Universidad de Salamanca, 37008 Salamanca, Spain.\\
$^2$ Instituto de F\'{\i}sica Corpuscular (IFIC), CSIC $\&$ Universitat de Valencia, Calle Catedr\'atico Jos\'e Beltr\'an, 2, 46980 Paterna, Valencia, Spain.\\
$^3$ Centro de L\'aseres Pulsados, CLPU, Parque Cient\'ifico, 37185 Villamayor, Salamanca, Spain.}

\begin{abstract}
The photon spectrum in macrocoherent atomic de-excitation via radiative emission of neutrino pairs (RENP) has been proposed as a sensitive probe of the neutrino mass spectrum, capable of competing with conventional neutrino experiments. In this paper we revisit this intriguing possibility, presenting an alternative method for inducing large coherence in a target based on adiabatic techniques. More concretely, we propose the use of a modified version of Coherent Population Return (CPR), namely double CPR, that turns out to be extremely robust with respect to the experimental parameters, and capable of inducing a coherence close to 100\,\% in the target.
\end{abstract}


\maketitle

\section{Introduction} \label{Introduction}

Neutrino oscillation experiments have demonstrated  that neutrinos are massive and there is leptonic flavour violation in their propagation \cite{Pontecorvo68, Gribov69} (see
\cite{GonzalezGarcia07} for an overview). Current data can be explained assuming that the three known neutrinos ($\nu_e$, $\nu_\mu$, $\nu_\tau$) are linear quantum superposition of three massive states $\nu_i$ ($i=1,2,3$) with masses $m_i$, connected by a Consequently, a leptonic mixing matrix which can be parametrized as \cite{PDG}:
\begin{equation}
  \label{eq:matrix}
  \rm U =
  \begin{pmatrix}
    \rm c_{12} c_{13}
    & \rm s_{12} c_{13}
    & 	\rm s_{13} e^{-i\delta_\text{CP}}
    \\
    \rm - s_{12} c_{23} - c_{12} s_{13} s_{23} e^{i\delta_\text{CP}}
    & \rm \hphantom{+} c_{12} c_{23} - s_{12} s_{13} s_{23}
    e^{i\delta_\text{CP}}
    &\rm  c_{13} s_{23}
    \\
    \rm \hphantom{+} s_{12} s_{23} - c_{12} s_{13} c_{23} e^{i\delta_\text{CP}}
    &\rm  - c_{12} s_{23} - s_{12} s_{13} c_{23} e^{i\delta_\text{CP}}
    & \rm c_{13} c_{23}
  \end{pmatrix} \begin{pmatrix}1 & 0 & 0 \\
0  & \rm e^{i\eta_1} & 0 \\ 
0  & 0 & \rm e^{i\eta_2} \end{pmatrix}
\end{equation}
where $\rm c_{ij} \equiv \cos\theta_{ij}$ and $\rm s_{ij} \equiv \sin\theta_{ij}$.  The phases $\eta_{i}$ are only non-zero if neutrinos are Majorana particles. If one chooses the convention where the angles $\theta_{ij}$ are taken to lie in the first quadrant, $\theta_{ij} \in [0, \pi/2]$,  and the CP phases $\delta_\text{CP},\eta_1,\eta_2 \in [0, 2\pi]$, then $\rm \Delta m^2_{21}=m_2^2-m_1^2>0$ by convention, and  $\rm \Delta m^2_{31}$ can be positive or negative. It is customary to refer to the first option as Normal Ordering (NO), and to the second one as Inverted Ordering (IO).

At present, the global analysis of neutrino oscillation data yields the three-sigma ranges for the mixing angles $\sin^2\theta_{12}$,  $\sin^2\theta_{23}$, and $\sin^2\theta_{13}$\cite{GonzalezGarcia14}, as well as for  the mass differences  $\rm \dfrac{\Dmq_{21}}{10^{-5}~\eVq}$ and $\rm \dfrac{\Dmq_{3\ell}}{10^{-3}~\eVq}$, but gives no information on the Majorana phases nor on the Dirac or Majorana nature of the neutrino. Also, neutrino oscillation experiments do not provide a measurement of the absolute neutrino masses, but only of their differences. Also, at present, oscillation experiments have not provided information on the mass ordering.  

The determination of the ordering and the CP violating phase $\delta_\text{CP}$ is the main goal of ongoing long baseline  (LBL) oscillation experiments \cite{Adamson14, Abe15, Patterson12}  which are sensitive to those in some part of the parameter space. Definite knowledge  is better guaranteed in future projects \cite{Abe15_2, Bass13}. 

Concerning the determination of the absolute mass scale in laboratory experiments, the standard approach is the search for the distortion of the end point of the electron spectrum in tritium beta decay. Currently, the most precise experiments \cite{Bonn01, Lobashev01} have given no indication in favour of distortion setting an upper limit 
\begin{equation}
\label{eq:nuelim}
\rm m_{\nu_e}=\left[\sum_i m^2_i |U_{ei}|^2\right]^{1/2} <2.2\,\text{eV} \; . 
\end{equation}
The ongoing KATRIN experiment \cite{Osipowicz01}, is expected to achieve an estimated sensitivity limit: $\rm m_\beta\sim0.3$\,eV.

The most precise probe of the nature of the neutrino is the search of neutrino-less double beta decay ($0\nu\beta\beta$) for verification of lepton number violation which is related to neutrino Majorana masses. So far, this decay has not been observed and the strongest bounds arise from experiments using $^{76}$Ge \cite{Macolino14}, $^{136}$Xe \cite{Gando13,Albert14}, and $^{130}$Te \cite{Alfonso15}. For the case in which the only effective lepton number violation at low energies is induced by the Majorana mass term for the neutrinos, the rate of $0\nu\beta\beta$ decay is proportional to the effective Majorana mass of $\nu_e$, and the experimental bounds on the corresponding lifetimes can be translated in constraints
on the combination  
\begin{equation}
\rm m_{ee}=\left| \sum_i m_i U_{ei}^2 \right|\lesssim 0.14\to 0.76 \; \text{eV}\, , 
\end{equation}
which, in addition to the masses and mixing parameters that affect the tritium beta decay spectrum, it also depends on two combinations of the CP violating phases $\delta_\text{CP}$ and $\eta_i$. 

A potentially revolutionary new way to explore fundamental neutrino physics may come from the field of quantum optics, thanks to recent technological advances. The key concept behind the intriguing possibility is the small energy difference between the levels in the atom or molecule, which allows for large relative effects associated with the small neutrino masses in the energy released in level transitions. This, in turn, opens  up the possibility of precision neutrino mass spectroscopy, as proposed by Ref.\,\cite{Yoshimura11, Fukumi12, Dinh12} and further explored by \cite{Song15}.

The relevant process in this case is the atomic de-excitation via radiative emission of neutrino pairs (RENP): $\rm |e\rangle \rightarrow |g\rangle + \gamma + \nu_i\bar \nu_j$.  The rate of this process can be made measurable if macro-coherence of the atomic target can be achieved \cite{Yoshimura12,Fukumi12}. The proposal is to reach such macro-coherent emission of radiative neutrino pairs via stimulation by irradiation of two trigger lasers of frequencies $\rm \omega, \omega'$ constrained by $\rm \omega + \omega' =\epsilon_{eg}/\hbar \,, \omega < \omega'$, with $\rm E_{eg} = E_e - E_g$ being the energy difference of initial and final levels. With this set-up the energy of the emitted photon in the de-excitation is given by the smaller laser frequency $\omega$ and therefore it can be very precisely known. Furthermore, neglecting atomic recoil, energy-momentum conservation implies that  each time the energy of the emitted photon  decreases below $\omega_{ij}$ with 
\begin{equation}
\rm \omega_{ij} = \frac{E_{eg}}{2} - \frac{(m_i +m_j)^2}{2E_{eg}}
\end{equation}
a new channel (this is, emission of another pair of massive neutrino species) is kinematically open. 

It is in principle possible to locate these thresholds energies by changing the laser frequency because the laser frequency, and therefore the emitted photon energy,  can be known with  high precision. Consequently, once the six $\omega_{ij}$ are measured, the spectrum of the neutrino masses could be fully identified. It has been argued that this method is ultimately capable of determining the neutrino mass scale, the mass ordering, the Dirac vs Majorana nature, as well as of  measuring the Majorana CP violating  phases \cite{Yoshimura11, Fukumi12, Dinh12}.

A necessary condition for RENP is macro-coherence. If the target is not prepared in a coherent state, the number of events per unit of time depends on the number N of particles of the target. This would be insufficient from an experimental point of view due to the extremely low cross section of the RENP process. However, if the target is in a coherent state the spectral rate scales with N$^2$. This situation is analogous to the process of superradiance suggested by Dicke in 1954 \cite{Dicke54}. According to this, one can conclude that one of the most critical aspects of this proposal is the development of a target as a robust coherent superposition of states. The original proposal of the Okayama group to generate the required coherent superposition was to induce Raman processes by irradiating the target with two different laser fields \cite{Yoshimura12_2}. In that case, two photon processes like $|\psi_3\rangle\leftrightarrow|\psi_1\rangle+\gamma+\gamma$ and $|\psi_3\rangle+\gamma\leftrightarrow|\psi_1\rangle+\gamma$  are driven populating state $|\psi_3\rangle$ and inducing therefore a coherence in the target.  However, it is important to consider that a Raman process is a second order process. This makes complicated to achieve a sufficient degree of control over the experimental parameters for inducing a large coherence in the medium. For example, Miyamoto \emph{et al} reported an induced coherence of $\sim$6.5\,$\%$ in Ref.\,\cite{Miyamoto14}.

In this manuscript we present an alternative method for inducing large coherence in a target based on adiabatic techniques. More concretely, we propose the use of a modified version of Coherent Population Return (CPR), namely double CPR, that turns out to be extremely robust with respect to the experimental parameters, and capable of inducing a coherence close to 100\,\% in the target. CPR belongs to the group of adiabatic techniques, sometimes also called coherent techniques, used to control the inherent coherent nature of laser-matter interaction. For example techniques like stimulated Raman adiabatic passage (STIRAP) \cite{Bergmann98} or electromagnetic induced transparency (EIT) \cite{Harris97, Marangos98} have proven to be a very valuable tool to steer population distributions, generate coherent superposition of quantum states, or support spectroscopical investigations. More concretely, CPR has been recently used  to prepare quantum superposition of states maximising therefore the nonlinear response of a medium \cite{Halfmann10, Chacon15}. In the following we will first briefly describe the basis of CPR. Then we will expand our study to the three-state system of interest for RENP. For the sake of an experimental implementation we will also study the robustness as well as the region of adiabaticity of the process with respect to the experimental parameters. Finally, we complete our contribution with a brief summary and an outlook.

\section{Coherent Population Return (CPR) technique}\label{CPR_Section}

Coherent Population Return is a coherent technique that produces a non-permanent transient of population of a target excited state. During the interaction with the laser radiation, it is possible to achieve a maximum population transfer of 50\,\%. Once the interaction ceases, the population returns adiabatically to the ground state. The original theoretical description of CPR can be found in \cite{Vitanov01}. From an experimental point of view, CPR has been successfully applied to the maintenance of the spectral resolution, i.e., to suppress power broadening at high laser intensities \cite{Vitanov01, Halfmann03, Peralta06}, in time-resolved femtosecond pump-probe experiments in molecular systems \cite{Peralta10}, and more recently to induced a robust coherent superposition of states for maximizing the nonlinear properties of a medium \cite{Halfmann10, Chacon15}.

Let us consider a two-state system with bare states $\{|\psi_1\rangle,|\psi_2\rangle\}$ interacting with a laser field whose frequency $\omega_{\rm L}$ is detuned from the Bohr frequency by $\Delta= {\rm E}_{21}-\omega_{\rm L} $, where ${\rm E_{21}}={\rm E_{2}}-{\rm E_{1}}$. As customary the description of the population evolution is based on the time-dependent Schr\"odinger equation with an appropriate interacting Hamiltonian:
\begin{equation}
\label{Schr}
i\hbar\frac{\partial{\rm |\Psi(t)\rangle}}{\partial t}=\rm H(t) |\Psi(t)\rangle.
\end{equation}
In the basis B formed by the states $\{|\psi_1\rangle,|\psi_2\rangle\}$, the Hamiltonian that describes this situation after  Rotating-Wave Approximation (RWA) \cite{Sh90} can be written as 
\begin{equation}
\label{hamiltonian}
\rm H(t)=\frac{\hbar}{2}\left(
\begin{array}{cc}
0 & \rm \Omega(t)\\
\rm \Omega(t) & 2\Delta
\end{array}
\right)
\end{equation}
where $\Omega$ is the Rabi frequency, i.e., the interaction energy divided by $\hbar$. Explicitly
\begin{equation}
\label{Rabi_def}
\Omega(\rm t)=\frac{\mu \rm E(t)}{\hbar}
\end{equation}
where $\mu$ is the electric dipole transition moment, and $\rm E(t)$ the electric field of the laser radiation.

If we define the statevector of the system $\rm |\Psi(t)\rangle$ in the basis B as
\begin{equation}
\rm |\Psi(t)\rangle=c_1(t)|\psi_1\rangle+c_2(t)|\psi_2\rangle,
\end{equation}
the Schr\"odinger equation (see Eq.\,\ref{Schr}) can be written as:
\begin{equation}
\label{Schr_matrix}
\rm \left(\begin{array}	{c} \dot{\rm c}_1(t) \\  \dot{\rm c}_2(t)\end{array}\right)=-\frac{\it{i}}{2}\left(
\begin{array}{cc}
0 & \rm \Omega(t)\\
\rm \Omega(t) & 2\Delta
\end{array}
\right)\left(\begin{array}{c}\rm c_1(t) \\ \rm c_2(t)\end{array}\right).
\end{equation}

The instantaneous basis B' $\{|\Phi_-\rm (t)\rangle, |\Phi_+\rm (t)\rangle\}$ that diagonalizes the Hamiltonian, i.e., in the adiabatic basis, can be written as 
\begin{equation}
\label{phiminus}
\begin{aligned}
 |\Phi_- \rm(t)\rangle=\cos \vartheta(t)|1\rangle- \sin \vartheta(t)|2\rangle\\
 |\Phi_+ \rm (t)\rangle=\sin \vartheta(t)|1\rangle+\cos\vartheta(t)|2\rangle,
\end{aligned}
\end{equation}
with associated eigenvalues $\left(\rm \lambda_+(t), \rm \lambda_-(t)\right)$
\begin{equation}
\label{lambdas}
 \rm \lambda_\mp(t)=\frac{1}{2}\left[\Delta\mp\sqrt{\rm\Omega^2(t)+\Delta^2}\right]
\end{equation}
where the mixing angle is defined by
\begin{equation}
\rm \vartheta(t)=(1/2)\arctan[\Omega(t)/\Delta].
\end{equation}
Accordingly, the eigenvector of the system in both basis can be the written as
\begin{equation}
\rm |\Psi(t)\rangle=c_1(t)|\psi_1\rangle+c_2(t)|\psi_2\rangle= c_-(t)|\Phi_- \rm(t)\rangle+c_+(t)|\Phi_+ \rm (t)\rangle,
\end{equation}
being the matrix that defines the basis change B'$\rightarrow$B defined by
\begin{equation}
\label{matrix_change}
\rm R\left[\rm \vartheta(t)\right]= \left(\begin{array}	{cc} \rm \cos \vartheta &  \rm \sin \vartheta \\
\rm -\sin \vartheta & \rm \cos \vartheta
\end{array}
\right).
\end{equation}
Taking into account that 
\begin{equation}
\rm |\Psi(t)\rangle_B=R\left[\rm \vartheta(t)\right]|\Psi(t)\rangle_{\rm B'}
\end{equation}
and 
\begin{equation}
\rm H_B(t)=R\left[\rm \vartheta(t)\right]H_{B'}(t)R^T\left[\rm \vartheta(t)\right]
\end{equation}
it is possible to express the Schr\"odinger equation of Eq.\,\ref{Schr} in the basis B':
\begin{equation}
\label{SchrBprime}
i\hbar\frac{\partial{\rm |\Psi(t)\rangle_{B'}}}{\partial t}=\rm \left(H_{B'}(t)-i\hbar R^T\left[\rm \vartheta(t)\right]\dot{R}\left[\rm \vartheta(t)\right]\right) |\Psi(t)\rangle_{B'}.
\end{equation}
According to the previous definition we can write Eq.\,\ref{SchrBprime} as
\begin{equation}
\label{Hamilt_adiab}
\rm \left(\begin{array}	{c} \dot{\rm c}_-(t) \\  \dot{\rm c}_+(t)\end{array}\right)=-\it{i}\left(
\begin{array}{cc}
\lambda_- & -i\dot{\vartheta}(t)\\
i\dot{\vartheta}(t) & \lambda_+
\end{array}
\right)\left(\begin{array}{c}\rm c_-(t) \\ \rm c_+(t)\end{array}\right).
\end{equation}
According to Eq.\,\ref{Hamilt_adiab} and in a situation where the non-diagonal terms are negligible with respect to the diagonal ones, if the statevector $\rm |\Psi(t)\rangle$ of the system is initially aligned with one of the adiabatic eigenstates, i.e., $\rm |\Phi_+(t)\rangle$ or $\rm |\Phi_-(t)\rangle$, it will remain parallel to it during the whole excitation process, i.e., the system will evolve adiabatically. Mathematically the adiabatic condition can be expressed as
\begin{equation}
\label{diff_eigenvectors}
\rm \lambda_+(t)-\lambda_-(t)\gg|\dot{\vartheta}(t)|,
\end{equation}
which turns out into the simple expression \cite{Vitanov01}:
\begin{equation}
\label{adiabatic_condition}
\rm |\Delta|\geq\frac{1}{T}
\end{equation}
being T the laser pulse duration. It is important to notice that the adiabatic region is exclusively determined by the laser detuning with respect to resonance and the duration of the laser pulse (or equivalently the laser bandwidth), being thus independent from the Rabi frequency $\rm \Omega(t)$. This makes CPR an extremely robust technique for its experimental implementation because these parameters are easily controllable in a real experiment.

\begin{figure}
\includegraphics[width=8.5cm]{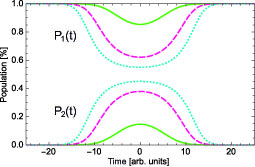}
\caption{\label{fig1} (Color online) Population of the excited and ground states versus time for different $\Omega_0/\Delta$ ratios. The Rabi frequency is $\Omega(t)=\Omega_0\exp{(-t^2/\tau^2)}$ with $\tau=8$ \,a.u. The green solid line corresponds to a $\rm \Omega_0/\Delta$ ratio of 1, the dashed pink line to $\rm \Omega_0/\Delta=4$, and the dotted blue line to $\rm \Omega_0/\Delta=10$.}
\end{figure}

Let us suppose now a situation where at the beginning of the interaction all the population starts in the ground (lower) state. Thus, we can write  $\rm |\Psi(t)\rangle=\rm |\Phi_-(t)\rangle=\rm |\psi_1\rangle$ since $t=-\infty\rightarrow\Omega(-\infty)=0\rightarrow\vartheta(-\infty)=0$. If the evolution is adiabatic, according to the previous analysis the statevector of the system  $\rm |\Psi(t)\rangle$ will remain always aligned to the adiabatic state  $\rm |\Phi_-(t)\rangle$. During the interaction, i.e., $-\infty<t<+\infty\rightarrow\Omega(t)\neq0\rightarrow\vartheta(t)\neq0$, $\rm |\Psi(t)\rangle$ is a coherent superposition of the bare states $|\psi_1\rangle$ and $|\psi_2\rangle$ defined by Eq.\,\ref{phiminus}. Once the interaction has ceased, i.e., $t=+\infty\rightarrow\Omega(+\infty)=0\rightarrow\vartheta(+\infty)=0$, the statevector of the system reads again $\rm |\Psi(+\infty)\rangle=\rm |\Phi_-(+\infty)\rangle=\rm |\psi_1\rangle$.  This means that the population transferred to the excited state during the process returns completely to the ground state once the interaction has ceased. As a consequence, no population resides permanently in the excited state no matter how large the transient intensity of the laser pump pulse might be (see Fig.\,\ref{fig1}).

It is interesting to notice that during the interaction the population of the excited states reads
\begin{equation}
\label{Popu_state2}
\rm P_2(t)=\sin^2\vartheta(t)=\frac{1}{2}-\frac{1}{2\sqrt{\left(\Omega(t)/\Delta\right)^2+1}}.
\end{equation}
Accordingly, in a situation where during the central part of the laser pulse $\rm \Omega(t)/\Delta\rightarrow\infty$, the population of the excited state will approach $\rm P_2(t)\rightarrow1/2$ obtaining therefore a perfect coherent superposition between ground and excited states in a robust way. At this point it is important to clarify the connection between macroscopic and microscopic quantities. The polarization of the sample (macroscopic quantity) is defined as $\rm P(t)=N\langle\Psi(t)|\mu|\Psi(t)\rangle$ being N the atomic density, $\rm \mu$ the electric dipole operator, and $\rm |\Psi(t)\rangle$ the eigenstate of the system. Taking into account the parity of the operator $\rm \mu$ it is possible to write $\rm P(t)\propto c_1(t)c_2^*(t)$ that is defined as the atomic coherence $\rm \rho_{12}(t)$ (microscopic quantity). The coherence, and hence the polarization, will reach a maximum in a situation where the population of ground and excited states are equal, i.e., $\rm |c_1(t)|^2=|c_2(t)|^2=1/2$. In other words, the polarization is maximum when the system is prepared in a perfect coherent superposition of states.  
 
Figure\, \ref{fig1} illustrates the facts discussed above. It shows the population of ground state $\rm P_1(t)$ and excited state $\rm P_2(t)$ for a fixed detuning $\Delta$ and different values of the Rabi frequency $\Omega_0$. The population of both states tends to the same value for high excitation intensities. In that case, and in the central part of the excitation pulse, the system is prepared in a perfect coherent superposition of the two states.

\section{Double CPR}
\begin{figure}
\includegraphics[width=8.5cm]{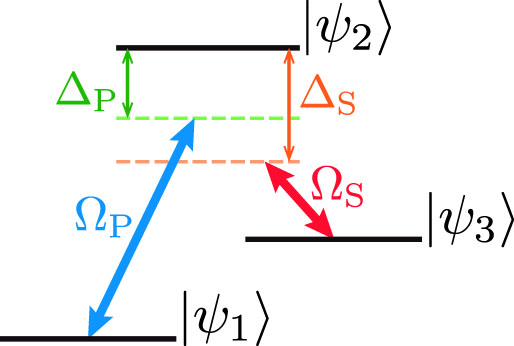}
\caption{\label{fig2} (Color online) Level scheme.}
\end{figure}

As it was mentioned in Section\,\ref{Introduction} the system of interest for the research of the neutrino hierarchy by means of laser-matter interaction techniques is a three-level system. In this system with states  $\{|\psi_1\rangle, |\psi_2\rangle,|\psi_3\rangle\}$, being $|\psi_2\rangle$ the state of maximum energy, the transitions between $|\psi_1\rangle\leftrightarrow|\psi_2\rangle$ and $|\psi_2\rangle\leftrightarrow|\psi_3\rangle$ are electric dipole allowed, while the transition $|\psi_1\rangle\leftrightarrow|\psi_3\rangle$ is forbidden. Consequently $|\psi_3\rangle$ is a metastable state (see Fig.\,\ref{fig2}). This system is often referred in the literature as a $\lambda$-system.

\subsection{Analytical analysis}
The hamiltonian that describes the situation depicted in Fig.\,\ref{fig2} after RWA can be written as:
\begin{equation}
\label{Hamilt_STIRAP}
\rm H(t)=\frac{\hbar}{2}\left(
\begin{array}{ccc}
0 & \rm \Omega_P(t) & 0\\
\rm \Omega_P(t) & \rm 2\Delta_P & \rm \Omega_S(t) \\
\rm 0 & \rm \Omega_S(t) &\rm 2(\Delta_P-\Delta_S)
\end{array}
\right),
\end{equation}
where $\rm \Delta_P$ and $\Delta_S$ are the detunings with respect to resonance of the lasers (Pump and Stokes) that couple the $|\psi_1\rangle\leftrightarrow|\psi_2\rangle$ and $|\psi_2\rangle\leftrightarrow|\psi_3\rangle$ transitions respectively. This Hamiltonian has been extensively studied for the situation of  Stimulated Raman Adiabatic Passage STIRAP (see for example \cite{Bergmann98} and references therein). In STIRAP, provided a two-photon resonance condition, i.e., $\rm \Delta_P=\Delta_S$, and a counterintuitive pulse sequence where the Stokes laser interacts with the system before the Pump laser, it is possible to achieve a complete and robust transfer of population from $|\psi_1\rangle$ to $|\psi_3\rangle$.  However, for obtaining a maximum coherent superposition between $|\psi_1\rangle$ and $|\psi_3\rangle$ it is necessary to equally distribute the population between both states. As it was explained in Section\,\ref{CPR_Section}, using CPR it is possible to obtain this superposition in a robust way in a two-level system. Thus, in this section we will extend that discussion to the more general situation depicted in Fig.\,\ref{fig2}.

\begin{figure}
\includegraphics[width=17cm]{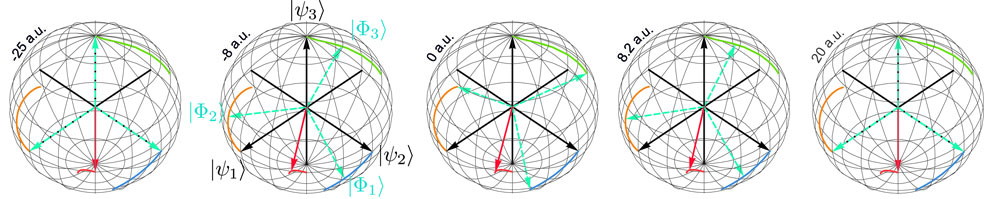}
\caption{\label{fig3} (Color online) Rotation of the adiabatic coordinate system defined by the $\rm |\Phi_i\rangle$} vectors (blue dotted lines) over the diabatic one defined by the $\rm |\psi_i\rangle$ vectors (black solid lines) as a function of time. The rotation axis (red solid line) as well as the different trajectories are also indicated.
\end{figure}

We will consider the most general situation where $\rm \Delta_P\neq\Delta_S$ and the following coefficients (we have dropped the time dependence for simplicity):
\begin{equation}
\begin{aligned}
\rm a=-\left(2\Delta_P-\Delta_S\right)\\
\rm b=\Delta_P\left(\Delta_P-\Delta_S\right)-\frac{1}{4}\left(\Omega_P^2+\Omega_S^2\right)\\
\rm c=\frac{1}{4}\left(\Delta_P-\Delta_S\right)\Omega_P^2.
\end{aligned}
\end{equation}
Thus, the eigenvalues of the Hamiltonian of Eq.\,\ref{Hamilt_STIRAP} read \cite{Sh90}:
\begin{equation}
\label{eigenvalues}
\begin{aligned}
\rm Z_1=-\frac{1}{3}a+\frac{2}{3}p\cos \frac{\theta}{3} \\
\rm Z_2=-\frac{1}{3}a-\frac{2}{3}p\cos\left(\frac{\theta}{3}+\frac{\pi}{3}\right)\\
\rm Z_3=-\frac{1}{3}a-\frac{2}{3}p\cos\left(\frac{\theta}{3}-\frac{\pi}{3}\right),
\end{aligned}
\end{equation}
where
\begin{equation} 
\label{equation_p}
\rm p=\sqrt{a^2-3b}
\end{equation}
and 
\begin{equation}
\label{angle_theta}
\rm \cos \theta=-\frac{27c+2a^3-9ab}{2p^3}.
\end{equation}
The basis B' that diagonalizes the Hamiltonian of Eq.\,\ref{Hamilt_STIRAP} can be written in terms of the vectors $\rm\{|w_1\rangle, |w_2\rangle, |w_3\rangle\}$ where
\begin{equation}
\label{ws}
\rm |w_i\rangle=\left(\Omega_P\left(Z_i-\left(\Delta_P-\Delta_S\right)\right), 2Z_i\left(Z_i-\left(\Delta_P-\Delta_S\right)\right), Z_i\Omega_S\right),
\end{equation}
for $\rm i=1,2, 3.$ Introducing the normalization factor 
\begin{equation}
\label{xis}
\rm \xi_i=\sqrt{\frac{Z_i}{\langle w_i|w_i\rangle\langle w_i|H|w_i\rangle}}
\end{equation}
the basis B' can be finally defined as $\rm\{|\Phi_1\rangle, |\Phi_2\rangle, |\Phi_3\rangle\}$, where
\begin{equation}
\label{Phis}
\rm |\Phi_i\rangle=\xi_i|w_i\rangle
\end{equation}
for  $\rm i=1,2, 3.$ Accordingly, the matrix that defines the basis change B'$\rightarrow$B is defined by
\begin{equation}
\label{matrix_change_STIRAP}
\rm R= \left(\begin{array}{ccc} \rm \xi_1\Omega_P\left(Z_1-\left(\Delta_P-\Delta_S\right)\right) &  \rm \xi_2\Omega_P\left(Z_2-\left(\Delta_P-\Delta_S\right)\right) & \rm \xi_3\Omega_P\left(Z_3-\left(\Delta_P-\Delta_S\right)\right)\\ \rm  \xi_12Z_1\left(Z_1-\left(\Delta_P-\Delta_S\right)\right) & \rm  \xi_22Z_2\left(Z_2-\left(\Delta_P-\Delta_S\right)\right) & \rm \xi_32Z_3\left(Z_3-\rm \left(\Delta_P-\Delta_S\right)\right)\\\rm \xi_1Z_1\Omega_S & \rm \xi_2Z_2\Omega_S & \rm \xi_3Z_3\Omega_S
\end{array}
\right).
\end{equation}
It must be noticed that for $\rm t=\pm\infty$, i.e., when both laser pulses are $\rm \Omega_{P, S}=0$, the vectors of the basis B' are aligned with the vectors of the basis B (see Appendix\,\ref{Projection}). More concretely, we have
\begin{equation}
\label{conexion}
\rm |\Phi_1\rangle=|\psi_2\rangle;\, |\Phi_2\rangle=|\psi_1\rangle;\, |\Phi_3\rangle=|\psi_3\rangle. 
\end{equation}
However, during the interaction, i.e., when $\rm \Omega_{P, S}\neq0$, the vectors of the basis B' can be expressed as a linear combination of the vectors of the basis B. Since by construction both basis are orthonormal, the new basis B' can be interpreted as a rotation of an angle over a certain axis of the original coordinate system defined by the vectors $\rm |\psi_i\rangle$ (see Fig.\,\ref{fig3}). The basis change B$\rightarrow$B', i.e., $\rm R^T$ according to the notation, can be written as:

\begin{equation}
\rm R^T =\left( \begin{array}{ccc} \rm \cos \alpha +u_x^2 \left(1-\cos \alpha\right) &\rm u_x u_y \left(1-\cos \alpha\right) - u_z \sin \alpha & \rm u_x u_z \left(1-\cos \alpha\right) + u_y \sin \alpha \\ \rm u_y u_x \left(1-\cos \alpha\right) + u_z \sin \alpha & \rm \cos \alpha + u_y^2\left(1-\cos \alpha\right) & \rm u_y u_z \left(1-\cos \alpha \right) - u_x \sin \alpha \\ \rm u_z u_x \left(1-\cos \alpha\right) - u_y \sin \alpha & \rm u_z u_y \left(1-\cos \alpha \right) + u_x \sin \alpha & \rm \cos \alpha + u_z^2\left(1-\cos \alpha\right) 
\end{array}
\right)
\end{equation}
where the rotation axis $\rm \textbf{u}=\left(u_x, u_y, u_z \right)$ must satisfy
\begin{equation}
\rm R^T\textbf{u}=\textbf{u},
\end{equation}
and the angle of rotation $\alpha$ is given by the following equation:
\begin{equation}
\label{alpha}
\rm \alpha=\arccos \frac{Tr\left(R^T\right)-1}{2},
\end{equation}
being $\rm Tr\left(R^T\right)$ the trace of the matrix $\rm R^T$.
In the Appendix\,\ref{Rotation} we show the expressions for the rotation axis $\rm \textbf{u}$ and the angle $\rm \alpha$.

Following the same argumentation of the previous section, for obtaining the adiabatic condition it is necessary to compare the Hamiltonian expressed in the base B', i.e., 
\begin{equation}
\label{Hamilt_STIRAP_B}
\rm H_{B'}=\left(
\begin{array}{ccc}
\rm Z_1 & 0  & 0\\
0 & \rm Z_2& 0 \\
0 & 0 &\rm Z_3
\end{array}
\right),
\end{equation}
with $\rm R^T\dot{R}$  (see Eq.\,\ref{diff_eigenvectors}). Explicitly $\rm R^T\dot{R}$ can be written as:
\begin{equation}
\label{no_adiabaticity}
\rm R^T\dot{R}=\left(
\begin{array}{ccc}
 0 &\rm a_{12}  &\rm a_{13}\\
\rm a_{21} & 0 &\rm a_{23} \\
 \rm a_{31} & \rm a_{32} & 0
 \end{array}
\right),
\end{equation}
with
\begin{equation}
\begin{aligned}
\rm a_{12}=-a_{21}=(\cos \alpha-1)(u_y\dot{u}_x-u_x\dot{u}_y)+\dot{u}_z\sin \alpha +\dot{\alpha}u_z\\
\rm a_{13}=-a_{31}= (\cos \alpha-1)(u_z\dot{u}_x-u_x\dot{u}_z)-\dot{u}_y\sin \alpha -\dot{\alpha}u_y\\
\rm a_{23}=-a_{32}=(\cos \alpha-1)(u_z\dot{u}_y-u_y\dot{u}_z)+\dot{u}_x\sin \alpha +\dot{\alpha}u_x.
\end{aligned}
\end{equation}
Since the diagonal elements $\rm a_{ii}$ are proportional to $\rm (u_x\dot{u}_x+u_y\dot{u}_y+u_z\dot{u}_z)$ and provided that $\textbf{u}$ is a unit vector, they result equal to zero. 

If at the beginning of the interaction the population is in the ground state, the statevector of the system is initially aligned with $\rm |\Phi_2\rangle$ (see Eq.\,\ref{conexion}). According to this, to investigate the conditions for adiabatic evolution the possible crossings between $\rm Z_2$ and $\rm Z_1$ or $\rm Z_3$ eigenenergies must be investigated. As it was discussed before, for avoiding the crossings between the eigenvectors, the difference between the eigenvalues must be larger than the non-diagonal terms that could produce this crossings. Mathematically, and taking into account the previous definitions, this can be expressed as (see Eq.\,\ref{eigenvalues}):
\begin{equation}
\label{Z_differences}
\begin{aligned}
\rm |Z_1-Z_2|=\left|p\left(\cos \frac{\theta}{3}-\frac{\sqrt{3}}{3}\sin\frac{\theta}{3}\right)\right|\gg\left|(\cos \alpha-1)(u_y\dot{u}_x-u_x\dot{u}_y)+\dot{u}_z\sin \alpha +\dot{\alpha}u_z\right |\\
\rm |Z_3-Z_2|=\left|\frac{2\sqrt{3}}{3}p\sin\frac{\theta}{3}\right| \gg\left|(\cos \alpha-1)(u_z\dot{u}_y-u_y\dot{u}_z)+\dot{u}_x\sin \alpha +\dot{\alpha}u_x\right |.
\end{aligned}
\end{equation}
Since p is a growing function with respect to the Rabi frequencies $\rm \Omega_P$ and $\rm \Omega_S$ (see Eq.\,\ref{equation_p}), the possible crossing will take place at the very beginning of the interaction. Thus, without loss of generality it is possible to set $\rm \Omega_{P,S}\rightarrow0$ in the previous equations obtaining
\begin{equation}
\label{Z_differences_2}
\begin{aligned}
\rm \left |p\left(\cos \frac{\theta}{3}-\frac{\sqrt{3}}{3}\sin\frac{\theta}{3}\right)\right |\gg 0\\
\rm \left |\frac{2\sqrt{3}}{3}p\sin\frac{\theta}{3}\right | \gg 0.
\end{aligned}
\end{equation}
According to this the conditions for the maximum and minimum of the left-hand side in Eq.\,\ref{Z_differences} can be shown to be (see Appendix\,\ref{Adiabatic_condition_appendix}):
\begin{equation}
\label{conditions}
\begin{aligned}
\text{minimum:}\, \rm \Delta_P=0\,\text{or}\,\Delta_P=\Delta_S\\
\rm |Z_1-Z_2|=|Z_3-Z_2|=0\\
\text{maximum:}\, \rm \Delta_P=2\Delta_S\,\Delta_S=2\Delta_P,\,\text{or}\,\Delta_P=-\Delta_S\\
\rm |Z_1-Z_2|=|Z_3-Z_2|=|\Delta_S|.\\
\end{aligned}
\end{equation}
The conditions of Eq.\,\ref{conditions} delimit the region of adiabaticity in the space parameters. It is worth noticing that similarly to the two-level system considered previously, the adiabatic condition for double-CPR does not depend on the Rabi frequency, being mainly determined by the detunings $\rm \Delta_P$ and $\rm \Delta_S$. 

\begin{figure}
\includegraphics[width=8.5cm]{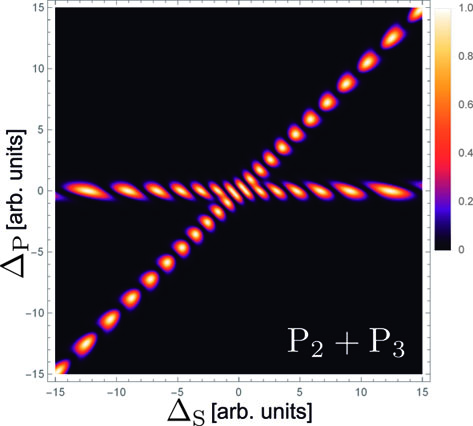}
\caption{\label{fig4} (Color online) $\rm P_2(+\infty)+P_3(+\infty)$ as a function of the detunings $\rm \Delta_P$ and $\rm \Delta_S$. The Rabi frequencies were $\rm \Omega_P=\Omega_S=\Omega_{0p}\exp{(-t^2/\tau^2)}$ with $\tau=6.5$\,a.u. and $\rm \Omega_{0p}=\Omega_{0s}=20\,a.u.$}
\end{figure}

Figure\,\ref{fig4} shows the sum of the population of states $|2\rangle$ and $|3\rangle$ at the end of the interaction for a $\rm \Omega_{P}=\Omega_{S}=20\,a.u.$ as a function of $\rm \Delta_P$ and $\rm  \Delta_S$ once the interaction has ceased ($\rm t=+\infty$). The numerical results of Fig.\,\ref{fig4} confirm the analytical adiabatic conditions of Eq.\,\ref{conditions} since in the perfect adiabatic case at the end of the interaction $\rm P_2=P_3=0$, i.e., at the end of the interaction all the population returns back to the ground state.   

\begin{figure}
\includegraphics[width=8.5cm]{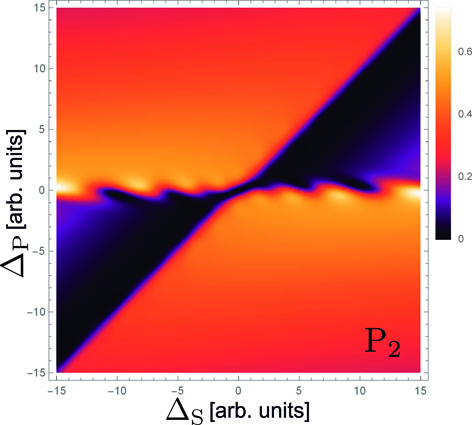}
\caption{\label{fig5} (Color online) $\rm P_2(0)$ as a function of the detunings $\rm \Delta_P$ and $\rm \Delta_S$. The Rabi frequencies were $\rm \Omega_P=\Omega_S=\Omega_{0p}\exp{(-t^2/\tau^2)}$ with $\tau=6.5$\,a.u. and $\rm \Omega_{0p}=\Omega_{0s}=20\,a.u.$}
\end{figure}

It is important to recognize that for a perfect induced coherence between states $|\psi_1\rangle$ and $|\psi_3\rangle$ it is necessary to fulfill the conditions for adiabaticity and also that no population resides in state $|\psi_2\rangle$ during the interaction.  As it was stated above if the conditions for adiabaticity are fulfilled, the state vector of the system remains always aligned with $|\Phi_2\rangle$. Figure\,\ref{fig5} shows the population of state $|\psi_2\rangle$ as a function of $\rm \Delta_P$ and $\rm \Delta_S$ for the maximum of the interaction (t=0). As it can be clearly seen, for the optimization of the process the detunings must be in the region defined by the conditions $\rm |\Delta_P|\geq\frac{1}{T}$ (adiabatic condition for single CPR) and $\rm \Delta_P=\Delta_S$.

\subsection{Numerical results}
\begin{figure}
\includegraphics[width=8.5cm]{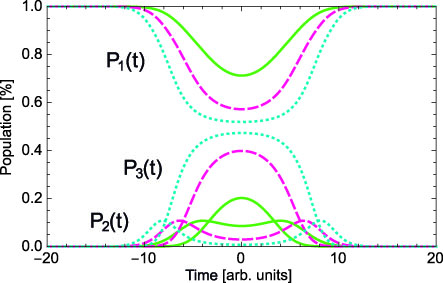}
\caption{\label{fig6} (Color online) Population of the different states as a function of time for different $\rm \Omega_{0p}/\Delta_P$ ratios. The Rabi frequencies were $\rm \Omega_P=\Omega_S=\Omega_{0p}\exp{(-t^2/\tau^2)}$ with $\tau=6$ \,a.u., and $\rm \Delta_S=2\Delta_P$. The solid green line corresponds to a $\rm \Omega_{0p}/\Delta_P$ ratio of 2, the dashed pink line to $\rm \Omega_{0p}/\Delta_P=4$, and the dotted blue line to $\rm \Omega_{0p}/\Delta_P=10$.}
\end{figure}

Figure\,\ref{fig6} shows the population dynamics for different $\rm \Omega_i/\Delta_P$ ratios provided that the detunings lie within the optimum region defined in the previous section. Similarly to the case of single CPR (see Fig.\,\ref{fig1}) the adiabaticity of the process is not influenced by the magnitude of the interaction, approaching $\rm P_1$ and $\rm P_3$ to 1/2 when the ratio of the Rabi frequencies over the detunings grows.  

\begin{figure}
\includegraphics[width=8.5cm]{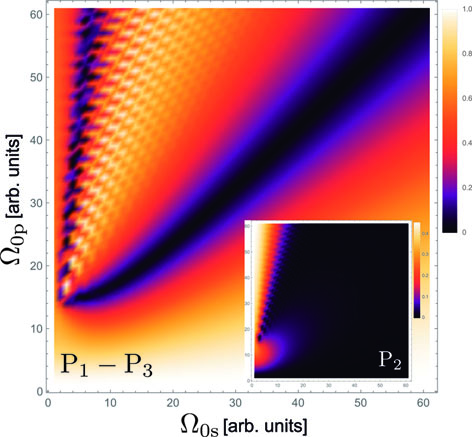}
\caption{\label{fig7} (Color online) $\rm P_1-P_3$ and $\rm P_2$ (inset) as a function of the Rabi frequencies for the maximum of the interaction (t=0). The simulation parameters were $\rm \Omega_P=\Omega_{0p}\exp{(-t^2/\tau^2)}$ and $\rm \Omega_S=\Omega_{0s}\exp{(-t^2/\tau^2)}$ with $\tau=6.5$ \,a.u and $\rm \Delta_S=2\Delta_P$ with $\rm \Delta_P=5\,a.u.$}
\end{figure}

For the experimental implementation of this technique it is mandatory to study the robustness of the process with respect to the laser intensity, i.e., with respect to the Rabi frequencies. Figure\,\ref{fig7} shows the difference $\rm P_1-P_3$ for a situation where $\rm P_2\rightarrow0$ as a function of $\rm \Omega_P$ and $\rm\Omega_S$. As it can be clearly seen the region of interest, i.e., the region where $\rm P_1-P_3\rightarrow0$ and $\rm P_2\rightarrow0$, is sufficiently large for assuring the experimental suitability of the technique. 

\section{Experimental considerations for Ba and Xe.}

In this section we will discuss some experimental issues regarding the implementation of a  macrocoherent state in barium and xenon. These are two of the possible candidates for the implementation of RENP \cite{Yoshimura12_2}. 

\subsection{Laser intensities}

In barium the level scheme is formed by the ground state $\rm 6s^2\,^1S_0$ as state $\rm |1\rangle$ (according to the notation used in this manusctript), state $\rm 6s6p\,^1P_1$ as $\rm |2\rangle$, and $\rm 6s5d\,^1D_2$ as $\rm |3\rangle$. The resonance between $\rm |1\rangle\leftrightarrow|2\rangle$ takes place at a wavelength of 553.7\,nm and with a transition moment of $\rm \mu_{12}=8\,D$ (1 Debye=3.33\,10$^{-30}$\,C$\cdot$m), and the resonance $\rm |2\rangle\leftrightarrow|3\rangle$ at 1500.4\,nm with $\rm \mu_{23}=0.2\,D$. Provided  detunings $\rm \Delta_P$ and $\rm \Delta_S$ lying in the adiabatic region defined by Fig\,\ref{fig4} and Fig.\,\ref{fig5}, Fig.\,\ref{fig7} determines the required Rabi frequencies. Assuming a pulse duration $\rm \tau=6.5\,ns$ (nanosecond lasers are normally used for adiabatic techniques because their bandwidths are of the same order of the natural bandwidths of the atomic states), and peak Rabi frequencies of $\rm \Omega_{0p}=\Omega_{0s}=20\,ns^{-1}$, and bearing in mind the relation between the Rabi frequency and the electric field (see Eq.\,\ref{Rabi_def}), and between the electric field and the intensity $\rm I(t)=\frac{1}{2}\epsilon_0 c E(t)^2$, the required peak intensities for the Pump and Stokes lasers are $\rm I_P\approx1\,kW cm^{-2}$ and  $\rm I_S\approx1.6\,MW cm^{-2}$ respectively.

For Xe the level scheme is more complicated. The ground state $\rm 5p^6\,^1S_0$ (state $|1\rangle$) is coupled via a two photon transition at a wavelength of 256\,nm to the state $\rm 5p^5(^2P_{3/2})6p^1\,^2\left[5/2\right]_2$ (state $|2\rangle$). The latter state is coupled via a one photon transition to the state $\rm 5p^5(^2P_{3/2})6s^1\,^2\left[3/2\right]_2$ (state $|3\rangle$) at a wavelength of 908\,nm. In the case of the two-photon transition between states $|1\rangle\leftrightarrow|2\rangle$ is necessary to calculate the effective two-photon Rabi frequency \cite{Sh90}:
\begin{equation}
\rm \Omega_P^{(2)}(t)=\sum_i \frac{\mu_{1i}\mu_{i2}}{2\hbar^2\Delta_{1i}}E^2(t). 
\end{equation}
This expression contains the transition moments from states $|1\rangle$ and $|2\rangle$ to all possible optically accessible intermediate states of the system, as well as the detuning of the pump laser to those states considering a one-photon transition. Using the data provided by \cite{Aymar78} we obtain a Rabi frequency $\rm \Omega_P^{(2)}\simeq0.1\,ns^{-1}$ per 1\,MWcm$^{-2}$. The Stokes transition $\rm |2\rangle\leftrightarrow|3\rangle$ has a transition dipole moment of $\rm \mu_{23}=5\,D$. According to this, for producing a Rabi frequency of 20\,$\rm ns^{-1}$ for both transitions it is required a pump laser intensity of $\approx$0.2\,GWcm$^{-2}$ and a probe intensity of $\approx$2\,kWcm$^{-2}$.

\subsection{Doppler shifts}

As we discussed in Section\,\ref{Introduction} the low cross section of the RENP process is compensated by the creation of a macrocoherent state scaling the spectral rate with N$^2$ being N the number of particles. According to this, the target particles must be contained in a close cell with the highest possible density. In this situation, the velocity of the particles as a consequence of the temperature of the cell defines different angles with respect to the laser beam producing a shift (Doppler shift) in the laser frequency as seen from the particle. This becomes specially relevant for CPR because the adiabatic region is exclusively defined by the detunings, i.e., by the difference between the Bohr frequency of the transition and the laser carrier frequency.  The magnitude of the Doppler shift is given by the equation:
\begin{equation}
\rm \Delta_{Doppler}=\overrightarrow{k}\cdot\overrightarrow{v}=\frac{2\pi v}{\lambda}\cos \phi.
\end{equation}
The velocity of the particles can be assumed to be in equilibrium and defined by:
\begin{equation}
\rm v=\sqrt{\frac{3kT}{m}}
\end{equation}
being k the Boltzmann constant and m the mass of the particle. The maximum Doppler shift is produced when the particle moves copropagating (or counterpropagating) with the laser beam ($\rm \phi=0,\pi$). Thus we can write: 
\begin{equation}
\rm \Delta_{Doppler}=\left|\frac{2\pi }{\lambda}\sqrt{\frac{3kT}{m}}\right|.
\end{equation}
As an example we can assume a detuning for the pump laser $\rm \Delta_P=10\,ns^{-1}$, and the transition $\rm |1\rangle \leftrightarrow |2\rangle$ at a wavelength of 553.7\,nm of Ba. For ensuring a robust adiabatic region it is necessary that $\rm \Delta_P\gg\Delta_{Doppler}$ resulting 
\begin{equation}
\rm T\ll4400\,K.
\end{equation}
Considering that the melting and boiling temperatures of Ba are 998\,K and 1413\,K respectively, the temperature obtained does not represent any additional experimental difficulty. 

\section{Conclusions}
As recently shown in \cite{Song15} the requirements for statistical determination of some crucial properties of the neutrino (in particular the neutrino mass scale and the mass ordering) using RENP are very demanding. In particular, one finds that,  even under ideal conditions, the determination of neutrino parameters needs experimental live time of the order of days to years for several laser frequencies, assuming  a target of volume of order 100\,cm$^3$ containing about $10^{21}$ atoms per cubic centimeter in a totally coherent state with maximum value of the electric field in the target. The technique presented in this paper may be an step forward towards attaining (and keeping) large coherence in macroscopic systems, e.g., $\sim10^{21}$ atoms per cubic centimetre. 

\section{Acknowledgements}
The authors thank N. Song and M.C. Gonzalez-Garcia for fruitful and stimulating discussions during the development of this manuscript. This work was supported by Ministerio de Econom\'ia y Competitividad, Spain (project FIS2014-53371-C4-3-R).

\appendix 
\section{Adiabatic basis projection}\label{Projection}
\begin{figure}
\includegraphics[width=8.5cm]{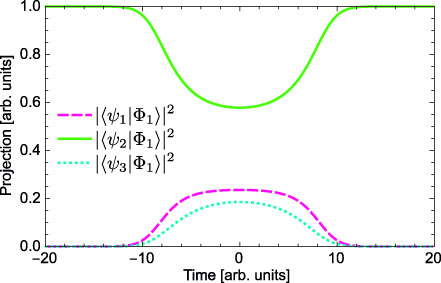}
\caption{\label{fig8} (Color online) $\rm |\Phi_1\rangle$ projection.}
\end{figure}

\begin{figure}
\includegraphics[width=8.5cm]{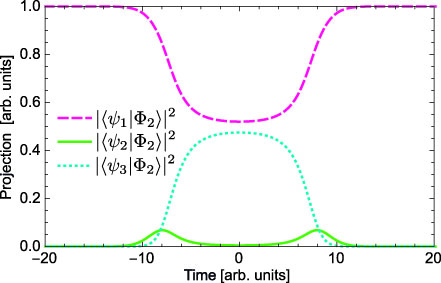}
\caption{\label{fig9} (Color online) $\rm |\Phi_2\rangle$ projection.}
\end{figure}

\begin{figure}
\includegraphics[width=8.5cm]{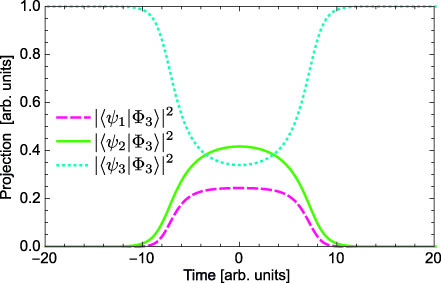}
\caption{\label{fig10} (Color online) $\rm |\Phi_3\rangle$ projection.}
\end{figure}

Figures\,\ref{fig8}, \ref{fig9} and \ref{fig10} show the square of the components of the vectors $\rm |\Phi_i\rangle$, i.e., the projection $\rm |\langle\psi_j|\Phi_i\rangle|^2$ for $\rm i,j=1,2,3$ as a function of time for a situation where $\rm \Omega_P=\Omega_S=\Omega_{0p}\exp{(-t^2/\tau^2)}$ with $\tau=6$ \,a.u.,  $\rm \Delta_S=(14/9)\Delta_P$, and  $\rm \Omega_{0p}/\Delta_P=50/9$. At the beginning of the interaction the adiabatic vector aligned with $\rm |\psi_1\rangle$ (assuming all the population starts in the ground state) is $\rm |\Phi_2\rangle$, being therefore the state vector of the system parallel to both. The statevector of the system will remain parallel to $\rm |\Phi_2\rangle$ if the evolution is adiabatic.

\section{Angle of rotation and axis.}\label{Rotation}
Defining the rotation axis $\rm \textbf{u}=(u_x, u_y, u_z)$ and with the help of the eigenvalues Z$_i$ (see Eq.\,\ref{eigenvalues}) and the normalization parameters $\xi_i$ (see Eq.\,\ref{xis}) the different components can be written as: 
\begin{eqnarray}
\rm u_x&=&\rm -\frac{a_1}{b_1 \sqrt{\frac{c_1+d_1+e_1+f_1}{g_1}}}\\
\rm a_1&=&\rm 2 Z_1 \xi_1(Z_1-\Delta_P+\Delta_S+Z_3 (-Z_1+Z_3) \xi_3 \Omega_S)\notag\\
\rm b_1&=&\rm -1+Z_3 \xi_3 \Omega_S+\xi_1 \Omega_P (Z_1-\Delta_P+\Delta_S-(Z_1-Z_3) (\Delta_S-\Delta_S) \xi_3 \Omega_S)\notag\\
\rm c_1&=&\rm 1+4 Z_1^2 (Z_1-\Delta_P+\Delta_S)^2 \xi_1^2+4Z_3^4 \xi_3^2-4 Z_3^2 (\Delta_P-\Delta_S) (2 Z_3-\Delta_P+\Delta_S) \xi_3^2\notag\\
\rm d_1&=&\rm Z_3 \xi_3 \Omega_S \left(-2-8Z_1^2 (Z_1-Z_3) (Z_1-\Delta_P+\Delta_S) \xi_1^2+Z_3 \left(1+4 Z_1^2 (Z_1-Z_3)^2\xi_1^2\right) \xi_3 \Omega_S\right)\notag\\
\rm e_1&=&\rm \xi_1^2 \Omega_P^2 ((Z_1-\Delta_P+\Delta_S)^2  (1+4 (Z_1-Z_3)^2 (Z_3-\Delta_P+\Delta_S)^2 \xi_3^2)\notag\\
&+&\rm (Z_1-Z_3) (\Delta_P-\Delta_S) \xi_3 \Omega_S (-2 (Z_1-\Delta_P+\Delta_S)+(Z_1-Z_3) (\Delta_P-\Delta_S) \xi_3 \Omega_S))\notag\\
\rm f_1&=&\rm 2 \xi_1 \Omega_P  ((Z_1-\Delta_P+\Delta_S) (-1+4 (Z_1-Z_3) Z_3 (Z_3-\Delta_P+\Delta_S)^2 \xi_3^2 )+\xi_3 \Omega_S (Z_1 (Z_3+\Delta_P-\Delta_S)+\notag\\
&+&\rm 2 Z_3 (-\Delta_P+\Delta_S)+Z_3 (-Z_1+Z_3) (\Delta_P-\Delta_S) \xi_3 \Omega_S))\notag\\
\rm g_1&=&\rm(1-Z_3\xi_3 \Omega_S+\xi_1 \Omega_P (-Z_1+\Delta_P-\Delta_S+(Z_1-Z_3) (\Delta_P-\Delta_S) \xi_3 \Omega_S))^2\notag
\end{eqnarray}

\begin{eqnarray}
\rm u_y&=&\rm \frac{(a_2+b_2)}{\sqrt{c_2-d_2-e_2+f_2}}\\
\rm a_2&=&\rm 1-Z_3 \xi_3 \Omega_S\notag\\
\rm b_2&=&\rm \xi_1 \Omega_P (-Z_1+\Delta_P-\Delta_S+(Z_1-Z_3) (\Delta_P-\Delta_S) \xi_3\Omega_S)\notag\\
\rm c_2&=&\rm 1+4 Z_1^2 (Z_1-\Delta_P+\Delta_S)^2 \xi_1^2+4 Z_3^4 \xi_3^2-4 Z_3^2(\Delta_P-\Delta_S) (2 Z_3-\Delta_P+\Delta_S) \xi_3^2\notag\\
\rm d_2&=&\rm Z_3 \xi_3 \Omega_S (-2-8 Z_1^2 (Z_1-Z_3) (Z_1-\Delta_P+\Delta_S) \xi_1^2+Z_3(1+4 Z_1^2 (Z_1-Z_3)^2\xi_1^2) \xi_3 \Omega_S)\notag\\
\rm e_2&=&\rm\ \xi_1^2 \Omega_P^2 ((Z_1-\Delta_P+\Delta_S)^2 (1+4 (Z_1-Z_3)^2 (Z_3-\Delta_P+\Delta_S)^2 \xi_3^2)+\notag\\
&+&\rm (Z_1-Z_3) (\Delta_P-\Delta_S) \xi_3 \Omega_S (-2 (Z_1-\Delta_P+\Delta_S)+(Z_1-Z_3) (\Delta_P-\Delta_S) \xi_3 \Omega_S))\notag\\
\rm f_2&=&\rm 2 \xi_1\Omega_P ((Z_1-\Delta_P+\Delta_S) (-1+4 (Z_1-Z_3) Z_3 (Z_3-\Delta_P+\Delta_S)^2 \xi_3^2)+\xi_3 \Omega_S (Z_1 (Z_3+\Delta_P-\Delta_S)+\notag\\
&+&\rm2 Z_3 (-\Delta_P+\Delta_S)+Z_3 (-Z_1+Z_3)(\Delta_P-\Delta_S) \xi_3 \Omega_S))\notag
\end{eqnarray}

\begin{eqnarray}
\rm u_z&=&\rm \frac{a_3}{b_3 \sqrt{\frac{c_3+d_3+e_3+f_3}{g_3}}}\\
\rm a_3&=&\rm 2 (Z_3-\Delta_P+\Delta_S) \xi_3 (Z_3+(Z_1-Z_3) (Z_1-\Delta_P+\Delta_S) \xi_1 \Omega_P)\notag \\
\rm b_3&=&\rm 1-Z_3 \xi_3 \Omega_S+\xi_1 \Omega_P (-Z_1+\Delta_P-\Delta_S+(Z_1-Z_3)(\Delta_P-\Delta_S) \xi_3 \Omega_S)\notag \\
\rm c_3&=&\rm 1+4 Z_1^2 (Z_1-\Delta_P+\Delta_S)^2 \xi_1^2+4 Z_3^4 \xi_3^2-4 Z_3^2 (\Delta_P-\Delta_S) (2 Z_3-\Delta_P+\Delta_S) \xi_3^2\notag \\
\rm d_3&=&\rm Z_3 \xi_3 \Omega_S (-2-8 Z_1^2 (Z_1-Z_3) (Z_1-\Delta_P+\Delta_S)\xi_1^2+Z_3 (1+4 Z_1^2 (Z_1-Z_3)^2 \xi_1^2) \xi_3 \Omega_S)\notag \\
\rm e_3&=&\rm \xi_1^2 \Omega_P^2 ((Z_1-\Delta_P+\Delta_S)^2 (1+4 (Z_1-Z_3)^2 (Z_3-\Delta_P+\Delta_S)^2 \xi_3^2)+\notag\\
&+&\rm (Z_1-Z_3) (\Delta_P-\Delta_S) \xi_3 \Omega_S (-2 (Z_1-\Delta_P+\Delta_S)+(Z_1-Z_3) (\Delta_P-\Delta_S) \xi_3 \Omega_S))\notag \\
\rm f_3&=&\rm 2 \xi_1 \Omega_P ((Z_1-\Delta_P+\Delta_S) (-1+4 (Z_1-Z_3) Z_3 (Z_3-\Delta_P+\Delta_S)^2 \xi_3^2)+\xi_3 \Omega_S (Z_1 (Z_3+\Delta_P-\Delta_S)\notag\\
&+&\rm 2 Z_3 (-\Delta_P+\Delta_S)+Z_3 (-Z_1+Z_3) (\Delta_P-\Delta_S) \xi_3 \Omega_S))\notag \\
\rm g_3&=&\rm (1-Z_3\xi_3 \Omega_S+\xi_1 \Omega_P (-Z_1+\Delta_P-\Delta_S+(Z_1-Z_3) (\Delta_P-\Delta_S) \xi_3 \Omega_S))^2\notag 
\end{eqnarray}

\begin{figure}
\includegraphics[width=8.5cm]{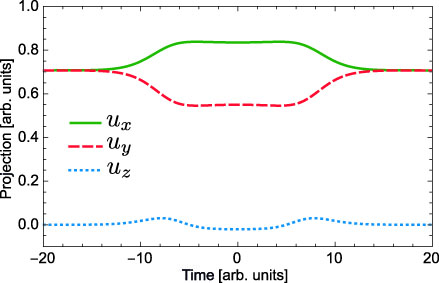}
\caption{\label{fig11} (Color online) Components $\rm u_x, u_y$ and $\rm u_z$ of the rotation axis as a function time for a situation where $\rm \Omega_P=\Omega_S=\Omega_{0p}\exp{(-t^2/\tau^2)}$ with $\tau=6$ \,a.u.,  $\rm \Delta_S=2\Delta_P$, and  $\rm \Omega_{0p}/\Delta_P=4$.}
\end{figure}

Figure\,\ref{fig11} shows the components $\rm u_x, u_y$ and $\rm u_z$ as a function of time. 

\begin{figure}
\includegraphics[width=8.5cm]{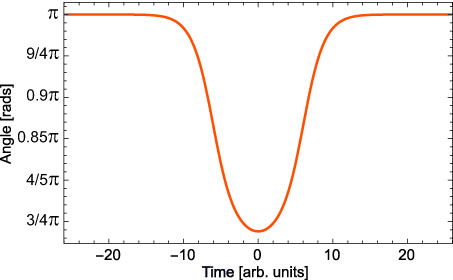}
\caption{\label{fig12} (Color online) Angle of rotation $\rm \alpha$ as a function of time for the same parameters that Fig.\,\ref{fig11}.}
\end{figure}

The rotation angle $\alpha$ (see Eq.\,\ref{alpha}) can be written as:
\begin{equation}
\rm 
\alpha=\arccos \frac{u_x^2-Z_1 \xi_1\Omega_P+(\Delta_P-\Delta_S) \xi_1 \Omega_P}{u_x^2-1}
\end{equation}
Figure\,\ref{fig12} shows the angle of rotation $\rm \alpha$ as a function of time.

\section{Adiabatic conditions}\label{Adiabatic_condition_appendix}

The maximum and minimum of the eigenvalues differences of Eq.\,\ref{Z_differences_2} are calculated as follows:
\begin{enumerate}
\item $\rm |Z_1-Z_2|=|p\left(\cos \frac{\theta}{3}-\frac{\sqrt{3}}{3}\sin\frac{\theta}{3}\right)|\\$
\begin{enumerate}

\item Minimum
\begin{equation}
\begin{aligned}
\rm \cos\left(\frac{\theta}{3}\right)-\frac{\sqrt{3}}{3} sin\left(\frac{\theta}{3}\right)=0\\
\rm \tan\frac{\theta}{3}=\sqrt{3}\\
\rm \theta_1= \pi+3n\pi\notag\\
\end{aligned}
\end{equation}

\item Maximum
\begin{equation}
\begin{aligned}
\rm \frac{\partial}{\partial\theta}\left(\cos \frac{\theta}{3}+\cos\left(\frac{\theta}{3}+\frac{\pi}{3}\right)\right)=0\\
\rm \sin\frac{\theta}{3}+\frac{\sqrt{3}}{3}\cos\frac{\theta}{3}=0\\
\rm \tan\frac{\theta}{3}=-\frac{\sqrt{3}}{3}\\
\rm \theta_2=-\frac{\pi}{2}+6n\pi
\end{aligned}
\end{equation}

\end{enumerate}

\item $\rm |Z_3-Z_2|=|\frac{2\sqrt{3}}{3}p\sin\frac{\theta}{3}|\\$
\begin{enumerate}

\item Minimum
\begin{equation}
\begin{aligned}
\rm \sin\frac{\theta}{3}=0\\
\rm \theta_3=3n\pi
\end{aligned}
\end{equation}

\item Maximum
\begin{equation}
\begin{aligned}
\rm \sin\frac{\theta}{3}=1\\
\rm \theta_4=\frac{3}{2}\pi+6n\pi
\end{aligned}
\end{equation}
\end{enumerate}
\end{enumerate}

Using the expression for the angle $\rm \theta$ (see Eq.\,\ref{angle_theta}) and choosing correctly the signs, we can determine the detunings for the maximum and minimum of $\rm |Z_i-Z_2|$.
\begin{enumerate}

\item Minimum
\begin{enumerate}

\item $\rm |Z_1-Z_2|$\\
For n=0 the equation 
\begin{equation}
\begin{aligned}
\rm \cos\theta_1=cos \pi=-1\\
\rm 2a^3-9ab-2p^3=0\\
\rm -2\Delta_P^3+3\Delta_P^2\Delta_S+3\Delta_P\Delta_S^2-2\Delta_S^3+2\left(\Delta_P^2-\Delta_P\Delta_S+\Delta_S^2\right)^{3/2}=0
\end{aligned}
\end{equation}
This last equation has two different solutions namely
\begin{equation}
\begin{aligned}
\rm \Delta_P=0\\
\rm \Delta_P=\Delta_S.
\end{aligned}
\end{equation}

\item $\rm |Z_3-Z_2|$\\
For n=0 the equation 
\begin{equation}
\begin{aligned}
\rm \cos\theta_3=cos\,0=1\\
\rm 2a^3-9ab+2p^3=0\\
\rm -2\Delta_P^3+3\Delta_P^2\Delta_S+3\Delta_P\Delta_S^2-2\Delta_S^3-2\left(\Delta_P^2-\Delta_P\Delta_S+\Delta_S^2\right)^{3/2}=0
\end{aligned}
\end{equation}
This equation has also two different solutions namely
\begin{equation}
\begin{aligned}
\rm \Delta_P=0\\
\rm \Delta_P=\Delta_S.
\end{aligned}
\end{equation}
\end{enumerate}

\item Maximum
\begin{enumerate}
\item $\rm |Z_1-Z_2|$\\
\begin{equation}
\rm \cos\theta_2=\cos-\frac{\pi}{2}=0
\end{equation}
that in terms of the detunings results
\begin{equation}
\begin{aligned}
\rm 2a^3-9ab=0\\
\rm \left(\Delta_P-2\Delta_S\right)\left(2\Delta_P-\Delta_S\right)\left(\Delta_P+\Delta_S\right)=0
\end{aligned}
\end{equation}
with solution
\begin{equation}
\begin{aligned}
\rm \Delta_P=2\Delta_S\\
\rm \Delta_S=2\Delta_P\\
\rm \Delta_P=-\Delta_S.
\end{aligned}
\end{equation}

\item $\rm |Z_3-Z_2|$\\
\begin{equation}
\rm \cos\theta_4=\cos\frac{3\pi}{2}=0
\end{equation}
with the same results that in the previous case
\begin{equation}
\begin{aligned}
\rm \Delta_P=2\Delta_S\\
\rm \Delta_S=2\Delta_P\\
\rm \Delta_P=-\Delta_S.
\end{aligned}
\end{equation}
\end{enumerate}
\end{enumerate}

\end{document}